\documentclass[a4paper,fleqn]{cas-sc}

\usepackage[authoryear,longnamesfirst]{natbib}

\def\tsc#1{\csdef{#1}{\textsc{\lowercase{#1}}\xspace}}
\tsc{WGM}
\tsc{QE}
\tsc{EP}
\tsc{PMS}
\tsc{BEC}
\tsc{DE}

\begin{document}
\let\WriteBookmarks\relax
\def\floatpagepagefraction{1}
\def\textpagefraction{.001}

\shorttitle{Integrating ChatGPT in a Computer Science Course: Students Perceptions and Suggestions}


\title [mode = title]{Integrating ChatGPT in a Computer Science Course: Students Perceptions and Suggestions}                      

\author[inst1]{Kehinde Aruleba }
\cormark[1]
\affiliation[inst1]{organization={School of Computing and Mathematical Sciences, University of Leicester},
           city={Leicester},
           postcode={LE1 7RH}, 
           country={UK, },
           email={ka388@leicester.ac.uk}}
\author[inst2]{Ismaila Temitayo Sanusi } \cormark[1]
\author[inst3]{George Obaido}
\author[inst4]{Blessing Ogbuokiri}

\affiliation[inst2]{organization={School of Computing, University of Eastern},
           city={Joensuu},
           postcode={P.O.Box 111, 80101}, 
           country={Finland },
           email={ismaila.sanusi@uef.fi}}

\affiliation[inst3]{organization={Center for Human-Compatible Artificial Intelligence (CHAI), Berkeley Institute for Data Science (BIDS), University of California},
           city={Berkeley},
           postcode={94720}, 
           state={California},
           country={USA },
           email={gobaido@berkeley.edu}}

\affiliation[inst4]{organization={Department of Mathematics and Statistics, York University},
           city={Toronto},
           postcode={ON M3J 1P3}, 
           country={Canada },
           email={blessogb@yorku.ca}}

\cortext[cor1]{Corresponding authors. ka388@leicester.ac.uk (K. Aruleba), ismaila.sanusi@uef.fi (I. Sanusi)}

\begin{abstract}
The integration of artificial intelligence tools such as ChatGPT in the education system has gained attention in recent years. This experience report explores students' perceptions and suggestions for integrating ChatGPT in a computer science course. Following a ChatGPT activity which includes code completion and analysis, seven students participated in in-depth interviews. Findings from the transcribed interviews suggest that ChatGPT has the potential to enhance learning experience including programming. They highlighted the tool's ability to respond immediately to queries and supporting personalised learning. However, they raise concerns that heavy reliance on ChatGPT may adversely affect students’ critical thinking and problem-solving skills. These findings show the importance of carefully balancing using ChatGPT in computer science courses. The findings of this research have significant implications for educators, curriculum designers and policymakers as they explore integrating AI tools into educational contexts.
\end{abstract}



\maketitle

\section{Introduction}\label{intr}\sloppypar
Educational institutions have undergone transformative shifts in recent years by embracing digital technology to support teaching and learning experiences \citep{aruleba2022evaluation, ayanwale2022teachers, aruleba2022technology,obaido2023analysis}. This shift is driven by various factors, including the rapid increase of online resources \citep{haleem2022understanding}, the need for more accessible and flexible learning experiences \citep{yunusa2021impact, criollo2021mobile}, and the need to equip students with the knowledge and skills needed for the 21st century \citep{agbo2021application, sanusi2022investigating}. Artificial intelligence (AI) systems have powered most of these shifts. AI-driven technologies and tools such as intelligent tutoring systems and recommendation systems all offer personalised experiences and support for students, administrators and educators \citep{xia2022self, sanusi2023investigating}. One area where these technologies have significantly impacted education is the development of chatbots including generative AI (GenAI) systems which has shown to support students learning \citep{FitsumEnhancing}. 

GenAI tools such ChatGPT are now becoming increasingly common, and they are being used across various sectors. Research has established ChatGPT’s capabilities to offer significant gains across different sectors including education \citep{dwivedi2023so, shoufan2023exploring}. The launch of ChatGPT which UNESCO \citep{UNESCO} tagged the fastest growing app in history popularized the GenAI concept among the general public. With ChatGPT’s distinct features and capability of discussing any topic of interest, it is currently one of the most powerful AI applications. As a chatbot that engage in convincing conversations with users, it is used to perform various tasks which includes essay writing, conducting literature reviews, enhancing papers, and writing computer code \citep{Owens}. \cite{Eva} asserts that the abilities of ChatGPT are expected to rapidly expand as it continues to receive new data through user interaction. These functionalities contribute to the contention of its role in education. 

While this AI tool could be used to enhance and support learning, its rapid growth and adoption raises integrity concerns such as cheating and plagiarism \citep{FitsumEnhancing}. Educators are mostly concerned about ChatGPT’s capabilities to produce an essay or coursework that will pass assessment or exams \citep{leinonen2023comparing}. While this concern is recognizable, the only way to address it is for educators to explore approaches that can be used to integrate the tool in classroom practices. As ChatGPT, a free and web-based tool which represents a tipping point in the development of AI \citep{Vaughan} come to stay in the educational system, we must figure out how to incorporate ChatGPT safely, effectively, and appropriately in teaching practices. Hence, this study is interested in how ChatGPT can be integrated into teaching and learning in a computer science (CS) course. We further explore the efficacy of using ChatGPT to learn programming in higher education institution (HEI). We recognise that past research has explored the possibility of learning various concepts including programming with ChatGPT \citep{leinonen2023comparing, ouh2023chatgpt}. However, more work is required, particularly that newer version of the GenAI platform keeps unfolding. More importantly is the fact that extensive research on the subject matter will assist education stakeholders to uncover strategies to effectively integrate ChatGPT in CS programs at the postsecondary level. 

This study is expected to provide insight into how ChatGPT can be incorporated into teaching and learning computing in HEIs classroom especially computing education. We anticipate that the outcome of this research will advance computer science education community’s knowledge among other field on the use of conversational chatbots in facilitating learning. This paper is organised as follows. Having highlighted the purpose of this study in Section~\ref{intr}, we discuss related work with respect to students perception of ChatGPT and its use for learning programming in Section~\ref{RW}. In Section~\ref{metho}, we describe our methods and activities given to participants and the follow-up interviews. In Section~\ref{find}, we present our key findings from the interview and the student activities. We discussed our findings and offer recommendations to relevant education stakeholders in Section~\ref{limit} and concluded with Limitations and future research directions. 

\section{Related works}\label{RW}
\subsection{Student perception of ChatGPT in higher education}
The emergence of ChatGPT holds profound implications for teaching and learning, particularly in HEIs. Studies examining the perceptions of stakeholders, especially students, within the context of various fields, including computer science, have been identified. \cite{shoufan2023exploring} conducted a two-stage analysis involving 56 senior students in a computer engineering program, revealing their appreciation for ChatGPT's capabilities in study and work. However, concerns about accuracy and the prerequisite background knowledge exist, emphasizing the necessity for educators to guide effective usage and advocate for improvements in model accuracy. Similarly, \cite{bonsu2023consumers} surveyed 107 students in Ghana, reporting positive perceptions and a willingness to adopt ChatGPT in education, despite no statistical relationship between perception and intention to use. \cite{firat2023chatgpt} explored ChatGPT's implications through the perspective of 21 scholars and students across four countries, identifying key themes such as the evolution of learning systems and ethical considerations. The study advocates for further research on ethics, privacy, and effective AI integration in education, concluding with an emphasis on understanding the opportunities and challenges of AI in higher education.

Additionally, a study by \cite{ngo2023perception} investigated how 200 university students perceive the use of ChatGPT for learning, revealing a generally positive view with benefits like time efficiency, diverse information access, personalized tutoring, and enhanced writing ideas. Concerns center around source evaluation, accurate citation, and language precision, proposing solutions such as verifying ChatGPT responses, utilizing it as a reference tool, providing guidelines, and fostering academic integrity to ensure ethical application in academia. \cite{hamid2023exploratory} explored the use of ChatGPT in process-driven problem-based learning (PDPBL) in the Bachelor of Pharmacy program. Results indicate that ChatGPT improves group collaboration, engagement, and motivation during PDPBL sessions. Despite challenges in understanding ChatGPT's information, most students recognize its potential to replace traditional information-seeking methods. The study suggests that ChatGPT can enhance PDPBL in pharmacy education, emphasizing the need for further research to validate the information provided and assess its impact on a larger sample size.

Meanwhile, \cite{elkhodr2023ict} investigated ChatGPT's role as a classroom aid in higher education, with three case studies involving ICT students. The findings reveal a positive perception of ChatGPT as a valuable learning resource, with students expressing a willingness to use AI tools in the future. The study suggests enhanced functionality, user flow, and content comprehension among ChatGPT users compared to those relying solely on traditional search engines. In \cite{woithe2023understanding}, the discussion of ChatGPT's dynamics in higher education, contribution, and implications was considered. Employing a constructivist-interpretivist approach with qualitative methods, the study uncovers factors influencing ChatGPT adoption, its post-adoption role, student and educator perspectives, and psychosocial effects on users. The trichotomous model underscores informed decision-making for integrating ChatGPT into education, balancing technical benefits with learner impacts. The study proposes UTAUT2 model extensions, prompting avenues for additional research in this evolving landscape.

\subsection{ChatGPT for teaching and learning programming}
In addition to studies exploring education stakeholders' perceptions of ChatGPT in language learning \cite{bin2023exploring, kohnke2023chatgpt}, there is a growing focus on the AI chatbot's role in supporting programming education. \cite{rahman2023chatgpt} delves into ChatGPT's impact on education and programming learning, examining personalized feedback and interactive teaching opportunities along with challenges like potential cheating and diminished critical thinking skills. The research includes coding experiments and surveys, providing insights into ChatGPT's benefits and challenges in programming education. In \cite{surameery2023use}, ChatGPT's use for debugging is explored, emphasizing its capabilities and limitations. The research recommends integrating ChatGPT into a broader debugging toolkit to optimize bug identification and resolution. \cite{opara2023chatgpt} investigated the introduction of ChatGPT in education, evaluating its benefits and challenges, including personalized learning experiences and task automation. Recommendations stress the importance of citing and referencing ChatGPT's responses in educational contexts.

\cite{ouh2023chatgpt} assess ChatGPT's efficacy in generating solutions for coding exercises in a Java programming course. Findings show its accuracy but note challenges with non-textual descriptions. \cite{leinonen2023comparing} evaluate ChatGPT's potential in producing code explanations, finding LLM-created explanations to be significantly easier to understand and more accurate than student-created ones. \cite{gottipati2023ai} address challenges in developing advanced programming skills, using ChatGPT as a support tool, showing initial evidence of its effectiveness. \cite{banic2023pair} explore pair programming with ChatGPT as an educational aiding tool, offering insights into this approach. \cite{chan2023students} survey students in Hong Kong, indicating enthusiasm for integrating ChatGPT in teaching, acknowledging its promise, and voicing concerns. \cite{yilmaz2023augmented} explores ChatGPT's impact on programming education, showing positive outcomes in computational thinking skills and motivation, and suggesting potential benefits for integration.

In summary, several studies indicate acceptance among students and teachers for the use or inclusion of ChatGPT as an aid to teaching and learning. This aligns with the purpose of this study, serving as an addition to knowledge with a distinct approach.

\section{Methodology}\label{metho}

This qualitative study aims to understand students' perception of integrating ChatGPT into the Computer Science curriculum. The study involved recruiting seven year 2 Computer Science students to complete some programming and database activities. After completing and submitting their attempts, they were then interviewed. All the interactions were done remotely due to the semester holidays. 

\subsection{Recruitment}
In this study, we employed quota sampling to select participants from the pool of year 2 Computer Science students. Our sample covers students from diverse demographics and academic abilities. The sample consisted of two female and five male participants, reflecting a gender-balanced representation within the cohort. Additionally, participants exhibited diversity in ethnic backgrounds, with one Black,  four Asian, and two White participants. The authors made an effort to ensure inclusivity by actively recruiting participants with disabilities. Two out of the seven participants in the sample reported having disabilities, enhancing the study's ability to explore the experiences and perspectives of students with diverse abilities within the context of computing education. All participants consented to participate in the study. As the study was conducted remotely, firstly,  we individually contacted and gave participants a brief about the research and shared questions and instructions about the activities.


\subsection{Procedure}
Due to the semester holiday, we developed a remote engagement protocol for the activities and interview (see Section~\ref{interv} and \ref{activities}). All the interaction and communications happened via email and Microsoft Teams call as it was the most convenient form of conversation for the participants. A brief email was sent to inquire about participants' availability for the study and describe the activities involved in the study and their expectations of the interview. In case of any concern, we provided participants the option to cancel or reschedule the call at any time during the study. The activities section of the study was divided into two parts (see Section~\ref{activities}), and the interview protocol was divided into five parts (see Section~\ref{interv}). Ethical approval was obtained before initiating the study. Informed consent was obtained from each participant, ensuring confidentiality, anonymity, and the right to withdraw at any stage.

\subsection{Data Collection and Analysis}
The data retrieved in this study were gathered through the students’ artefacts, completed learning tasks, and cognitive interviews. Participants were firstly told to submit screenshots of each of their codes alongside the code generated by ChatGPT, including the prompts given to ChatGPT. Except for where the authors already provided Java code snippets, all the activities and feedback submitted by students were written in different programming languages due to the instruction that allows students to use a language of their choice.

Secondly, we followed up with participants a week after submitting the answers to the activities for interview requests. We conducted Microsoft Teams based structured interviews about ChatGPT. The interview duration ranges from 18 - 30 minutes. In total, we collected seven individual feedbacks on the activities and 184 minutes of interview data. The interviews were recorded after obtaining consent and later transcribed. Subsequently, qualitative coding and analysis were done. Based on the grounded theory approach, codes were developed based on emergent themes from the interviews \citep{constantinou2017comparative}. These codes were refined through several coding iterations until theoretical saturation was achieved. A simultaneous literature review process also informed the thematic analysis. Two authors of this study were responsible for conducting all qualitative coding and analysis. 

Throughout the process, a colleague provided iterative feedback on the coding and analysis. The initial step involved reviewing our transcribed data several times to internalise the various interviewee views and accounts. All authors spent several weeks interacting with the data, which helped to establish their credibility. The authors combined the overlapping codes and eliminated the duplicates before carefully selecting the final codes. All of the authors actively participated in several rounds of peer debriefing \citep{houghton2013rigour} to settle significant differences that arose during this procedure.

\subsection{ChatGPT Activities}\label{activities}
To understand the possibilities of integrating ChatGPT in a Computer Science course, we designed activities with clear instructions for participants. These tasks were distributed to participants via email during a holiday period, allowing for flexibility in completion time. Participants were told to attempt the questions independently and also use ChatGPT to solve the questions after attempting them; this will be useful for code comparison, i.e. comparing human code vs ChatGPT-generated code. They were further instructed to take screenshots of their code and that of ChatGPT. The activities were divided into two categories, i.e., code completion and code analysis.

\subsubsection{Code Completion}
Participants were given a series of code completion tasks covering various programming languages and concepts, including user input/output, SQL queries, and array manipulation. This task assessed participants' ability to complete code snippets and solve programming challenges. Some tasks require them to write the code from the beginning, and others are partial code snippets with missing sections, and participants are required to solve the problem by completing the code. The questions were carefully selected to cover a range of complexity levels, i.e., from basic to intermediate. Each participant was presented with a total of five code completion tasks. Participants were explicitly informed not to get help or support from any external source (such as Internet and textbooks) when attempting the tasks. However, they were encouraged to think critically and attempt all questions independently. 

\subsubsection{Code analysis}
Participants were given two codes and instructed to analyse them based on specific criteria. The criteria included identifying potential bugs, suggesting code optimisations and solutions, and explaining the overall functionality of the code. The code snippets were written in Java and administered individually to each participant. The tasks included analysing output in a given string manipulation code and identifying and addressing potential issues in a given calculator class, including error handling and data loss concerns.

Responses from participants, including code snippets and explanations, were collected via a secure cloud platform. The data included screenshots of ChatGPT and participants' code solutions and PDF documents containing their explanations and analyses. No personal identifying information was collected to ensure anonymity.

\subsection{Participants and Interview}\label{interv}
This study was conducted in the context of a Computer Science course. We specifically engage our participants with programming and database activities for the study duration. Students were interviewed with a semi-structured interview procedure. These students have all completed and passed at least a programming and database course in the past; this is expected to provide more indepth insights into how ChatGPT can be used in coding, hence contributing additional critical perspective. The interview questions were divided into five different categories. This was done to enhance the quality and effectiveness of the research process, and also to systematically explore different aspects of the participants' experiences and opinions. The categorisation aimed to provide structure to the interview process and maintain focus during the interview. The categories are defined as follows:

\begin{itemize}

\item Activities Questions: Participants were asked to compare their code with ChatGPT-generated code, focusing on structure, readability, and syntax. Creativity and innovation in the AI-generated code were explored, along with strengths and weaknesses in both.

\item General Questions: The broader implications of ChatGPT in education were discussed. Participants shared their perspectives on how ChatGPT could assist educators, its integration into existing educational platforms, support for students with disabilities, potential drawbacks, and the necessary preparation for ChatGPT classroom use.

\item Ethical Questions: Participants reflected on the ethical considerations surrounding ChatGPT, expressing their views on its impact on teaching, learning, and society. Opinions on students' use of ChatGPT for learning tasks were also discussed.

\item Evaluating Questions (after the activities): Participants provided their overall impressions of ChatGPT and discussed any concerns regarding its ethical implications that may have arisen during the comparison activities.

\item Wrap-up Questions: The interview concluded with participants sharing insights gained from the study and their plans for applying these insights to their coding practices or future projects. 

\end{itemize}

The interviews were conducted online by one of the article's authors, who has prior experience teaching a programming course to the participants. This approach aimed to establish a level of familiarity and trust between the interviewer and participants, creating an environment conducive to open and honest responses. The interviewer's teaching background also provided valuable context for understanding the participants' perspectives within the context of a programming course. However, efforts were made to minimize potential biases by maintaining a neutral stance during the interview. The interviews were scheduled and conducted individually to ensure focused and undivided attention. 

\section{Findings}\label{find}
Four themes were generated from the qualitative coding and analysis. These themes were further divided into sub-themes. The themes are student perception on using ChatGPT for coding, student suggestions in incorporating ChatGPT in teaching and learning, ethics and student general perception. 

\subsection{Student Perception on using ChatGPT for coding}
Despite our interview, students perceived a range of benefits and drawbacks in using ChatGPT for coding. Some of these are explained below. 

\subsubsection{Code structure}
We asked participants if they noticed any differences between the structure of their code and that of ChatGPT; most participants described the structure of ChatGPT-generated codes as straightforward and easy to follow, even for novice programmers. P6, for example, said, “\textit{in terms of readability, ChatGPT would like split it up. So, for example, I remember the first question when it was taking first and last name; it did first name and last name on different lines and it made it more concise question for the user to answer}.” Similarly, P1 said, “\textit{in one of the tasks, I have created an extra variable to do the calculations. However, ChatGPT generated code combined it in one variable, and I am shocked; this made the code easier and shorter than mine}.” 

Consistent with findings from \cite{khoury2023secure}, one of the participants noted that ChatGPT codes can be “\textit{generic and insure}” (P3), and there is not much creativity in there. Despite this comment from P3, other participants, such as P1 and P6 believed thet ChatGPT’s code could be innovative and improved over time, showing the potential for ChatGPT to introduce creative approaches to coding tasks. Overall, the students’ perceptions of using ChatGPT for their coding activities were generally positive. These findings suggest that this tool is promising and could assist and enhance coding tasks while also challenging students to think creatively about problem-solving.

\subsubsection{Code comment and explanation}
\begin{figure}[t]  
\centering
    \includegraphics[width=10cm]{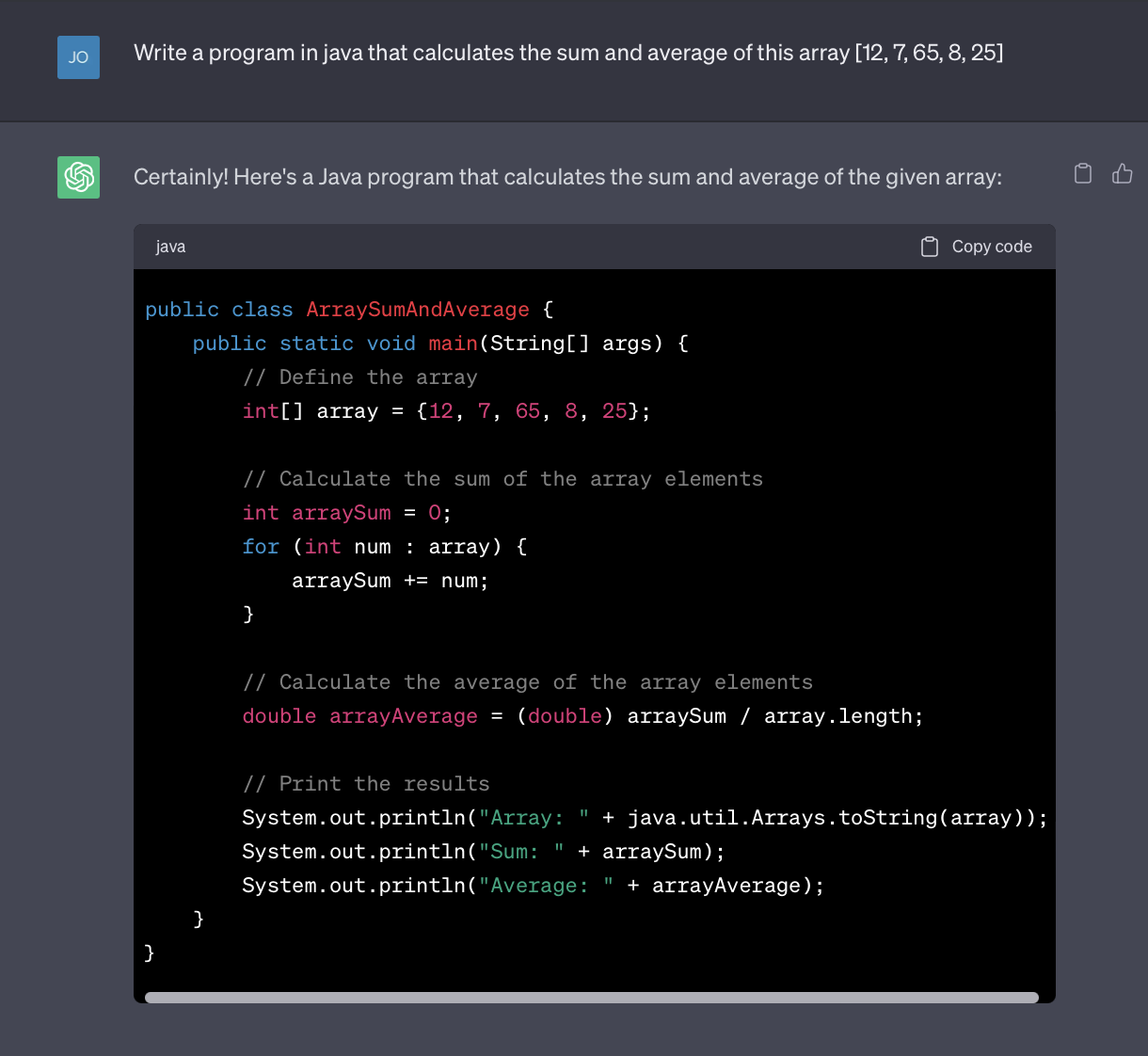}
    \caption{P6 - ChatGPT generated code with comments}
    \label{fig:image1}
\end{figure}

The majority of the participants reported that ChatGPT-generated codes are well commented and believed that this can potentially improve student programming skills. P1 stated that “ChatGPT provided detailed comments in a straightforward approach.” Multiple participants also shared this sentiment, indicating that ChatGPT’s code comments are generally perceived as helpful in learning.
More specifically, participants emphasised and appreciated that these comments were generated automatically without prompting ChatGPT to add comments to the code (see Figure~\ref{fig:image1}). Five participants highlighted this feature of ChatGPT. P1 noted that even though they are a “\textit{very good programmer, I rarely add comments to my code even when I am asked to; I just add two or three words as comments}.”

The result suggested mixed reactions when asked if they think ChatGPT default comments could be better if they prompted it to add comments to the code. P3 said that could lead to “\textit{confusion for the user because ChatGPT comments explain even the simplest things (variables), which makes the comments too much}.” Similarly, P5 also raised concerns about the accuracy of the comments \textit{“if ChatGPT is instructed to add comment, it will go too in-depth and might not be correct at the end}". These concerns indicate that while comprehensive comments can be helpful, they must also be accurate to maintain meaning. P7 noted that despite their practice of adding comments to their code, they believed that their comments might not match ChatGPT standards. This indicates that some participants perceived ChatGPT comments to be of higher quality than their own. The differences between user and AI-generated comments show the importance of aligning the AI’s commenting technique with user code practices and preferences. 

Some participants, such as P4 perceived ChatGPT’s comments as more than just explanations – they viewed them as educational tools that could facilitate and support learning programming. P3, P6, P7 shared a similar view - they believe the comments are like explanations and descriptive enough, which will help them learn programming quickly. ChatGPT’s comments have the potential to not only help in code understanding but also facilitate the learning of programming concepts.

\subsubsection{Self-learning of programming} \sloppypar
Our findings showed that all participants agreed that tools like ChatGPT can significantly enhance self-learning in programming. P4 stated, "\textit{ChatGPT's ability to instantly respond to their queries has been instrumental in their learning journey}.” Four participants highlighted the tool's potential to ease the challenges encountered by students who struggle with programming. They emphasised that ChatGPT can make the lives of such students easier when used appropriately. P7 noted that the feature of ChatGPT to simplify complex concepts and real-time assistance is valuable for those encountering programming difficulties. Similarly to some of the findings in \citet{yilmaz2023augmented}, P2 said they had effectively used ChatGPT to “\textit{accelerate their programming learning journey}.” According to P7, ChatGPT has the ability to empower students in their creative pursuits, “\textit{it is a tool that offers guidance, helping students tap into their creativity and innovation}.”

\subsection{Student suggestions in incorporating ChatGPT in teaching and learning}
All the participants highlighted that ChatGPT is a highly effective tool for supporting their learning process. More specifically, they noted that ChatGPT has the potential to assist lecturers in the development of learning materials and assessments.

\subsubsection{Teaching}
The findings revealed suggestions from participants on how ChatGPT can be used to help in teaching. All participants pointed out that lecturers can use ChatGPT to generate various activities for students based on their strengths. P1 said, “\textit{… during my internship as a teacher for A-level students, I used ChatGPT to generate tasks and lab questions for some of the lessons, which was quite helpful}.” P4 suggested that schools should offer a course for students on how to use ChatGPT and its implications. This education and training would ensure that students can harness the tool's full potential.

Similarly, P5 said, “\textit{… universities can create resources to support our lecturers effectively using ChatGPT – training, guidelines or best practices.}” Some participants suggested using ChatGPT in the classroom as an alternative to Courseware Platforms will provide more interactive and adaptive learning experiences. According to the participants, the tool can be integrated into learning management systems (LMS) such as Blackboard and Canvas; they echoed that adding it to the online forum section of LMS will greatly help lecturers. P6 highlighted that “\textit{when you ask questions in online forums, you either wait for a fellow student or the lecturer to respond. I believe integrating ChatGPT will be beneficial because it answers you immediately, and you do not have to wait for hours}.”

\subsubsection{Learning}
All the participants stated that ChatGPT has the potential to help students learning process; this is consistent with the findings in  \cite{shoufan2023exploring}. A common theme among participants was ChatGPT's diverse problem-solving approaches. They expressed their excitement about the tool's ability to solve a single question in multiple approaches; these approaches enhance problem-solving and critical thinking skills by exposing students to different approaches and methods. P7 noted that “\textit{… I can use the regenerate button several times to have multiple versions of a question. Reading these different versions of answers to a single question will broaden my knowledge}.” They all found ChatGPT to be helpful in gaining a deeper understanding of the learning material. According to P2, “\textit{I prefer chatGPT to access reading resources than traditional textbooks provided by lecturers.}” This shift towards interactive and digital materials is seen as a more flexible and engaging learning method for most participants. 

\begin{figure}[t]  
\centering
    \includegraphics[width=10cm]{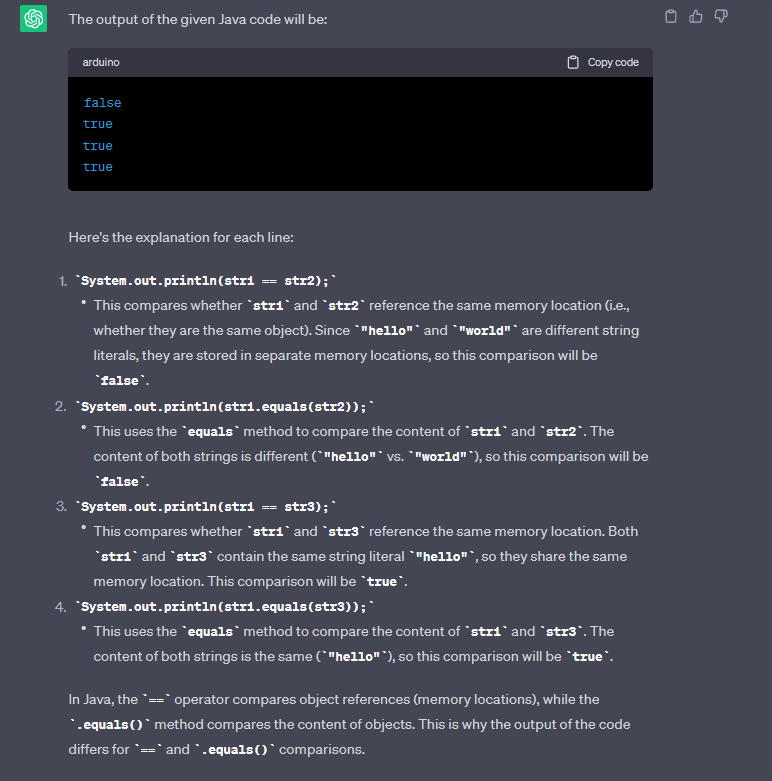}
    \caption{P4 - ChatGPT explaining the output of a code}
    \label{fig:image3}
\end{figure}

ChatGPT could supplement lecturers' explanations (P2, P3, P7). Due to the limited time allocated for lectures, these participants said they used the tool to further understand concepts that were not explained in detail in class (see Figure~\ref{fig:image3}). P1 also highlighted efficient code debugging, “\textit{… working on an assignment and having bugs or errors that would take hours to go through Stackoverflow and all of these websites, ChatGPT will just help me out and really explain what the problem was}.” Despite the positives about ChatGPT, P3 raised concerns about the accuracy of the tool responses. They feel the responses might sometimes be “\textit{… outdated and incorrect information}”. This suggests that the responses should be carefully reviewed and evaluated for assessments or learning tasks.

\subsubsection{Student support}
\begin{figure}[t]  
\centering
    \includegraphics[width=10cm]{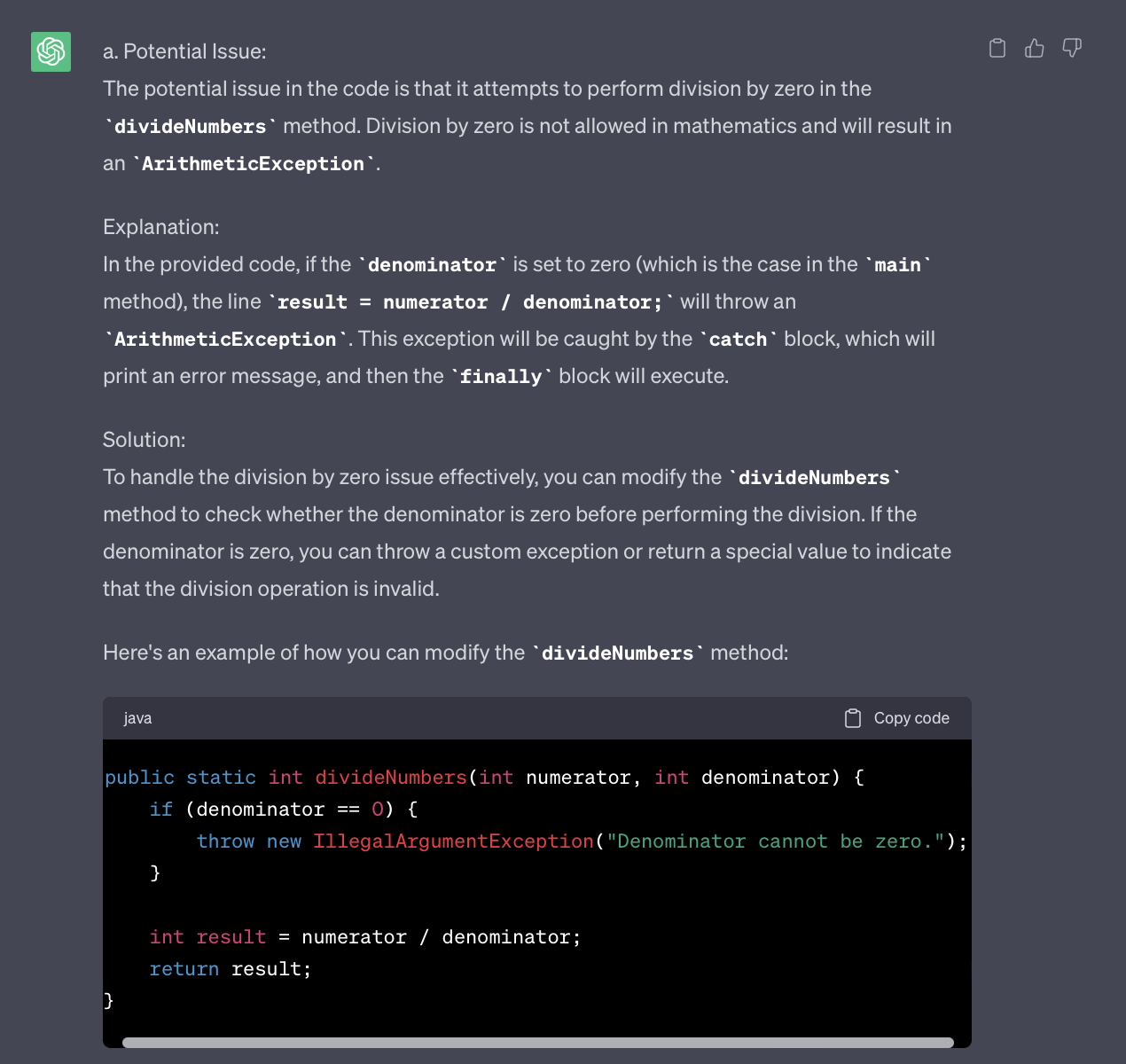}
    \caption{P2 - Error identification and possible solutions explained by ChatGPT}
    \label{fig:image2}
\end{figure}

Most participants said they now prefer to get help from ChatGPT instead of search engines because they may find the solution to the errors or bugs in their code using search engines but do not understand how this solution works. However, with ChatGPT, it explains the errors and bugs as shown in Figure~\ref{fig:image2}. P2 said, “\textit{… it is valuable because you do not need to go through all Python documentation to know exactly what functions you need to use}”. Consistent with  \cite{fuchs2023exploring}, P2 emphasises the importance of having an “\textit{… individual/customised personalised feedback mechanism}” and the significance of having access to an AI-based tutor system they can approach whenever they need help (P4). All participants agreed that the tool will effectively support students with disabilities. P7 notably stated that it will support those with dyslexia, and the tool's flexibility in writing styles will be beneficial in addressing students’ specific disability-related needs. 

P3 acknowledge they have used ChatGPT in “\textit{… writing emails and grammar checks}.” This feature was said to support students learning to speak and write in English potentially. P1, P2, P3, and P7 stated that ChatGPT explains code effectively, and this has stopped them from copying and pasting code they know little about from the internet. Some of the participants raised concerns. P1 and P3 said that when the question is not constructed properly, the responses from ChatGPT “\textit{… confused them.}” Similar to \cite{sok2023chatgpt}, all participants mentioned concerns about students becoming overreliant on the tool. This will reduce students' critical thinking and problem-solving skills. P2 and P6 stated that overreliance on ChatGPT could lead to reduced attention during lectures and encourage students to think less when solving problems. It is essential that integrating ChatGPT in educational systems should aim to have a balance between providing personalised support and promoting self-reliance.


\subsection{Ethics}
All the study participants are concerned about the ethical implications of using ChatGPT in the teaching and learning process. While plagiarism appears to be the prominent issue which resonates with the concerns of different stakeholders, especially in the academic \citep{anders2023using} settings, other issues related to privacy, data protection and misinformation were raised.

\subsubsection{Plagiarism} Consistent with earlier concerns in literature \citep{anders2023using, cotton2023chatting}, the students reported copying and pasting of content directly from the chatbot constitute a challenge for effective usage of the AI for learning. The participants believe the AI app is now being used by students to generate resources and they take credit for work that is not theirs. For instance, P4 stated that "\textit{one of the ethical implications is students taking credit for work that is not theirs, which is a very bad practise in and outside of academic institutions.}" P5 also opined that "\textit{...some people might claim ChatGPT output as their work.}" With the concerns about cheating in mind, P4 suggested that "\textit{educational institutions should offer a short course on ChatGPT usage and as part of the training, students should be prompted on how they would feel claiming credit for work that is not theirs.}"

\subsubsection{Privacy} The study participants identified privacy as a major concern. This concern is linked to the storage of the users data on the AI app. As stated by P2, "\textit{using chatGPT raises concerns because all of the users data is stored in it, whatever you type in the chatbot.}" Perhaps some of the participants are not aware that new ways to manage personal data in ChatGPT now exist \cite{OpenAI}. According to  \cite{OpenAI} it is possible as from April 2023 that end users’ to turn off chat history so that their data won’t be used to train the Chatbot models by default. 

\subsubsection{Data protection} Similar to privacy concern, two participants specifically mentioned data protection. In what appears to be two sides of same coin with privacy principles, data protection to the participants is how to limit ChatGPT's collection and retention of personal information. P3 expressed his concern by mentioning that "\textit{... if the user passes a university data, that will go directly into the database and anyone will be able to get that information.}"

\subsection{Student general perception}
From the participants interview data, we identified their general perception about ChatGPT which we categorized as strengths and weaknesses. 

\subsubsection{ChatGPT strengths} With regards to the AI chatbot strength, the participants highlighted several positive impact of its use. The notable take-aways from the students responses include: ChatGPT's capability to support personal study plan, assisting struggling learners, and usefulness for English language writing, especially non-native speakers. 

According to p2, "\textit{... is very effective for personal study time because AI can provide things that are personal to you and it is also available for you all the time.}" Some of the participants hold the believe that it assist in clarifying lessons taught after classes. P3 opine that "\textit{ChatGPT will be a good follow-up tool after classroom sessions to enhance learning, especially struggling learners.}" In line with existing research that ChatGPT promotes English language learning \citep{bin2023exploring, kohnke2023chatgpt}, P5 stated that ChatGPT "\textit{is pretty accurate especially in correcting grammatical mistakes and [he] usually uses it to try to see if mistake was made or not.}"

Overall, the participants rated the AI chatbot as "\textit{time saver unlike web searching,} p4" with the capability to "\textit{give real-time feedback to users prompt and customised learning materials}" P6. There is also a general consensus that ChatGPT is a valuable tool for learning and improving ones knowledge. 

\subsubsection{ChatGPT weaknesses} Despite the strengths attributed to ChatGPT usage and its effectiveness in improving learning, the students have concerns regarding heavy reliance on the tool. The participants prominent considerations include: lessening of critical thinking \& problem solving capacity, negative effect on students creativity and responsibility, and misinformation. 

Some participants expressed the opinion that the constant usage of ChatGPT to complete task or exercises may limit learners problem solving skills. For instance P2 stated that "\textit{....just looking up code on ChatGPT and like getting examples from that instead of you know trying out the code yourself would lessen your critical thinking and your problem solving capacity.}" Relatedly, P4 reason that "\textit{relying on ChatGPT heavily will hinder the individual's creativity and innovation such that students will not be able to come up with fully authentic solutions or attempt difficult problems.}" P6 also believe that continual usage will impact significantly on ones responsibility. The participant specifically described what he meant by responsibility in this context as "\textit{the sense that you [users] are not responsible for it [the content] if ChatGPT generated it.}" Participant 1 \& 3's responses suggested that without information verification, misinformation can set in based on heavy reliance on ChatGPT. 

\section{Discussion}\label{disc}
In recent years, the integration of artificial intelligence and natural language processing technologies into educational settings has gained momentum. One significant application of these technologies is the incorporation of chatbots, such as ChatGPT, into the curriculum of computer science courses. In this study, our aim was to explore students' perceptions and gather their suggestions regarding the integration of ChatGPT as an instructional tool in a computer science course. We specifically investigated students' perceptions and suggestions for integrating the GenAI tool into computer science courses following ChatGPT coding activities. Through interviews, we gathered the thoughts of seven participants involved in our study. The conversations with the study participants focused on their experiences with coding using ChatGPT, ethical concerns, suggestions for integrating ChatGPT into their courses, and their general opinions about its usage. Our findings suggest that students believe ChatGPT is a promising tool for learning, including programming. Additionally, students suggested adopting a well-defined model for the incorporation of ChatGPT into their courses. Furthermore, we identified from the students' feedback that heavy reliance on ChatGPT may diminish critical thinking and problem-solving skills. The ethical considerations raised by the participants include plagiarism, privacy, and data protection.

Our findings align with previous studies that have examined the use of ChatGPT in educational settings. Similar to the positive perceptions reported in studies exploring the use of ChatGPT for student support and engagement \citep{shoufan2023exploring, bernabei2023students}, our participants appreciated the convenience and accessibility of ChatGPT. This consistency suggests that ChatGPT can serve as a valuable resource to enhance programming experiences by providing timely and accessible assistance outside of traditional classroom hours. ChatGPT's availability for on-demand support contributes to fostering autonomous learners, particularly in the context of programming. Our students recognized ChatGPT's role in enhancing their self-learning experiences within programming. This finding aligns with the broader literature on self-directed learning \citep{yilmaz2023augmented, baskara2023promises}, highlighting the potential of ChatGPT to empower learners to take ownership of their educational journeys and foster a more independent and self-driven approach to programming education. Another advantage identified in our study was ChatGPT's utility in assisting with code restructuring. This aligns with previous research on automated code refactoring tools, emphasizing the potential for ChatGPT systems to aid programmers in optimizing and enhancing code quality \citep{cao2023study}. ChatGPT's contributions to code restructuring align with the broader goals of improving code maintainability and readability.

However, our study also uncovered challenges and concerns identified in prior research related to the integration of GenAI in education. The inconsistencies in ChatGPT's responses, as highlighted by our participants, have been a common issue reported in previous studies \citep{joshi2023chatgpt, currie2023academic}. These inconsistencies underscore the need for ongoing refinement and optimization of ChatGPT responses to ensure accuracy and reliability, particularly in educational contexts where the quality of information is critical. Such inconsistencies may reinforce misinformation, which can be counterproductive to learning because students are less likely to identify these errors. Our study revealed a potential challenge associated with the integration of ChatGPT: the potential reduction of critical thinking among students. This concern is consistent with research on the potential drawbacks of excessive reliance on GenAI in education \citep{shoufan2023exploring}. Overreliance on ChatGPT for code-related tasks may inadvertently hinder students' abilities to engage in deep critical thinking and problem-solving, essential skills in computer science. Such overreliance may negatively impact learning and lead to academic dishonesty \citep{kasneci2023chatgpt}. This concern highlights the importance of carefully designed integration strategies that strike a balance between the use of AI tools and the traditional teaching methods employed by educators.

It is crucial to acknowledge the limitation related to data privacy concerns. While ChatGPT serves as a valuable educational tool, its operation may involve the collection and retention of personal information. This limitation aligns with the broader discourse on data privacy and security in ChatGPT \citep{wu2023unveiling}. Striking a balance between providing personalized learning experiences and safeguarding the privacy of students' data is a paramount challenge in the integration of ChatGPT in education. This limitation necessitates the implementation of robust data protection measures, ensuring compliance with relevant regulations such as the GDPR in Europe or the Family Educational Rights and Privacy Act (FERPA) in the United States. Educators and institutions must adopt stringent policies and practices that limit ChatGPT's collection and retention of personal information to only what is essential for its educational functions. Moreover, transparent communication with students about data usage and privacy safeguards is imperative to build trust and maintain ethical standards in AI-enhanced education.

In light of our participants' feedback and suggestions, our study contributes to the ongoing discourse surrounding the role of AI in education. Recommendations for customization, personalization, increased interactivity, and transparency align with the evolving landscape of GenAI-enhanced educational technologies \citep{leinonen2023comparing}. These suggestions not only reflect the desire for more tailored and engaging learning experiences but also emphasize the importance of clear communication and guidance regarding the role and limitations of ChatGPT in the educational process..




\section{Limitation and Future work}\label{limit}

We identified some limitations in our study. First, the coding activities that our participants completed were not complex as reflected in their responses to our interviews. Designing complex task may generate more insight. Second, this study did not consider the ChatGPT version adopted by each participants to complete the exercises. Future research may explore how different versions of ChatGPT or other GenAI tools perform in completing coding tasks. Third, we worked with only seven students in this study which account for the use of a qualitative method. Future research may conduct similar study with relatively large sample and analyzing the outputs from human vs AI.

\section{Acknowledgement}
 We thank the students who participated in this study.



\bibliographystyle{elsarticle-harv}

\bibliography{sample-base}

\begin{thebibliography}{50}
\expandafter\ifx\csname natexlab\endcsname\relax\def\natexlab#1{#1}\fi
\providecommand{\url}[1]{\texttt{#1}}
\providecommand{\href}[2]{#2}
\providecommand{\path}[1]{#1}
\providecommand{\DOIprefix}{doi:}
\providecommand{\ArXivprefix}{arXiv:}
\providecommand{\URLprefix}{URL: }
\providecommand{\Pubmedprefix}{pmid:}
\providecommand{\doi}[1]{\href{http://dx.doi.org/#1}{\path{#1}}}
\providecommand{\Pubmed}[1]{\href{pmid:#1}{\path{#1}}}
\providecommand{\bibinfo}[2]{#2}
\ifx\xfnm\relax \def\xfnm[#1]{\unskip,\space#1}\fi
\bibitem[{Agbo et~al.(2021)Agbo, Sanusi, Oyelere and
  Suhonen}]{agbo2021application}
\bibinfo{author}{Agbo, F.J.}, \bibinfo{author}{Sanusi, I.T.},
  \bibinfo{author}{Oyelere, S.S.}, \bibinfo{author}{Suhonen, J.},
  \bibinfo{year}{2021}.
\newblock \bibinfo{title}{Application of virtual reality in computer science
  education: A systemic review based on bibliometric and content analysis
  methods}.
\newblock \bibinfo{journal}{Education Sciences} \bibinfo{volume}{11},
  \bibinfo{pages}{142}.
\bibitem[{Anders(2023)}]{anders2023using}
\bibinfo{author}{Anders, B.A.}, \bibinfo{year}{2023}.
\newblock \bibinfo{title}{Is using chatgpt cheating, plagiarism, both, neither,
  or forward thinking?}
\newblock \bibinfo{journal}{Patterns} \bibinfo{volume}{4}.
\bibitem[{Aruleba et~al.(2022a)Aruleba, Jere and
  Matarirano}]{aruleba2022evaluation}
\bibinfo{author}{Aruleba, K.}, \bibinfo{author}{Jere, N.},
  \bibinfo{author}{Matarirano, O.}, \bibinfo{year}{2022}a.
\newblock \bibinfo{title}{An evaluation of technology adoption during remote
  teaching and learning at tertiary institution by gender}.
\newblock \bibinfo{journal}{IEEE Transactions on Computational Social Systems}
  .
\bibitem[{Aruleba et~al.(2022b)Aruleba, Jere and
  Matarirano}]{aruleba2022technology}
\bibinfo{author}{Aruleba, K.}, \bibinfo{author}{Jere, N.},
  \bibinfo{author}{Matarirano, O.}, \bibinfo{year}{2022}b.
\newblock \bibinfo{title}{Technology adoption readiness in disadvantaged
  universities during covid-19 pandemic in south africa.}
\newblock \bibinfo{journal}{International Journal of Higher Education}
  \bibinfo{volume}{11}, \bibinfo{pages}{172--180}.
\bibitem[{Ayanwale et~al.(2022)Ayanwale, Sanusi, Adelana, Aruleba and
  Oyelere}]{ayanwale2022teachers}
\bibinfo{author}{Ayanwale, M.A.}, \bibinfo{author}{Sanusi, I.T.},
  \bibinfo{author}{Adelana, O.P.}, \bibinfo{author}{Aruleba, K.D.},
  \bibinfo{author}{Oyelere, S.S.}, \bibinfo{year}{2022}.
\newblock \bibinfo{title}{Teachers’ readiness and intention to teach
  artificial intelligence in schools}.
\newblock \bibinfo{journal}{Computers and Education: Artificial Intelligence}
  \bibinfo{volume}{3}, \bibinfo{pages}{100099}.
\bibitem[{Bani{\'c} et~al.(2023)Bani{\'c}, Konecki and Konecki}]{banic2023pair}
\bibinfo{author}{Bani{\'c}, B.}, \bibinfo{author}{Konecki, M.},
  \bibinfo{author}{Konecki, M.}, \bibinfo{year}{2023}.
\newblock \bibinfo{title}{Pair programming education aided by chatgpt}, in:
  \bibinfo{booktitle}{2023 46th MIPRO ICT and Electronics Convention (MIPRO)},
  \bibinfo{organization}{IEEE}. pp. \bibinfo{pages}{911--915}.
\bibitem[{Baskara(2023)}]{baskara2023promises}
\bibinfo{author}{Baskara, F.R.}, \bibinfo{year}{2023}.
\newblock \bibinfo{title}{The promises and pitfalls of using chat gpt for
  self-determined learning in higher education: An argumentative review}, in:
  \bibinfo{booktitle}{Prosiding Seminar Nasional Fakultas Tarbiyah dan Ilmu
  Keguruan IAIM Sinjai}, pp. \bibinfo{pages}{95--101}.
\bibitem[{Bernabei et~al.(2023)Bernabei, Colabianchi, Falegnami and
  Costantino}]{bernabei2023students}
\bibinfo{author}{Bernabei, M.}, \bibinfo{author}{Colabianchi, S.},
  \bibinfo{author}{Falegnami, A.}, \bibinfo{author}{Costantino, F.},
  \bibinfo{year}{2023}.
\newblock \bibinfo{title}{Students’ use of large language models in
  engineering education: A case study on technology acceptance, perceptions,
  efficacy, and detection chances}.
\newblock \bibinfo{journal}{Computers and Education: Artificial Intelligence} ,
  \bibinfo{pages}{100172}.
\bibitem[{Bin-Hady et~al.(2023)Bin-Hady, Al-Kadi, Hazaea and
  Ali}]{bin2023exploring}
\bibinfo{author}{Bin-Hady, W.R.A.}, \bibinfo{author}{Al-Kadi, A.},
  \bibinfo{author}{Hazaea, A.}, \bibinfo{author}{Ali, J.K.M.},
  \bibinfo{year}{2023}.
\newblock \bibinfo{title}{Exploring the dimensions of chatgpt in english
  language learning: A global perspective}.
\newblock \bibinfo{journal}{Library Hi Tech} .
\bibitem[{Bonsu and Baffour-Koduah(2023)}]{bonsu2023consumers}
\bibinfo{author}{Bonsu, E.}, \bibinfo{author}{Baffour-Koduah, D.},
  \bibinfo{year}{2023}.
\newblock \bibinfo{title}{From the consumers’ side: Determining students’
  perception and intention to use chatgptin ghanaian higher education}.
\newblock \bibinfo{journal}{Available at SSRN 4387107} .
\bibitem[{Cao et~al.(2023)Cao, Li, Wen and Cheung}]{cao2023study}
\bibinfo{author}{Cao, J.}, \bibinfo{author}{Li, M.}, \bibinfo{author}{Wen, M.},
  \bibinfo{author}{Cheung, S.c.}, \bibinfo{year}{2023}.
\newblock \bibinfo{title}{A study on prompt design, advantages and limitations
  of chatgpt for deep learning program repair}.
\newblock \bibinfo{journal}{arXiv preprint arXiv:2304.08191} .
\bibitem[{Chan and Hu(2023)}]{chan2023students}
\bibinfo{author}{Chan, C.K.Y.}, \bibinfo{author}{Hu, W.}, \bibinfo{year}{2023}.
\newblock \bibinfo{title}{Students' voices on generative ai: Perceptions,
  benefits, and challenges in higher education}.
\newblock \bibinfo{journal}{arXiv preprint arXiv:2305.00290} .
\bibitem[{Constantinou et~al.(2017)Constantinou, Georgiou and
  Perdikogianni}]{constantinou2017comparative}
\bibinfo{author}{Constantinou, C.S.}, \bibinfo{author}{Georgiou, M.},
  \bibinfo{author}{Perdikogianni, M.}, \bibinfo{year}{2017}.
\newblock \bibinfo{title}{A comparative method for themes saturation (comets)
  in qualitative interviews}.
\newblock \bibinfo{journal}{Qualitative research} \bibinfo{volume}{17},
  \bibinfo{pages}{571--588}.
\bibitem[{Cotton et~al.(2023)Cotton, Cotton and Shipway}]{cotton2023chatting}
\bibinfo{author}{Cotton, D.R.}, \bibinfo{author}{Cotton, P.A.},
  \bibinfo{author}{Shipway, J.R.}, \bibinfo{year}{2023}.
\newblock \bibinfo{title}{Chatting and cheating: Ensuring academic integrity in
  the era of chatgpt}.
\newblock \bibinfo{journal}{Innovations in Education and Teaching
  International} , \bibinfo{pages}{1--12}.
\bibitem[{Criollo-C et~al.(2021)Criollo-C, Guerrero-Arias,
  Jaramillo-Alc{\'a}zar and Luj{\'a}n-Mora}]{criollo2021mobile}
\bibinfo{author}{Criollo-C, S.}, \bibinfo{author}{Guerrero-Arias, A.},
  \bibinfo{author}{Jaramillo-Alc{\'a}zar, {\'A}.},
  \bibinfo{author}{Luj{\'a}n-Mora, S.}, \bibinfo{year}{2021}.
\newblock \bibinfo{title}{Mobile learning technologies for education: Benefits
  and pending issues}.
\newblock \bibinfo{journal}{Applied Sciences} \bibinfo{volume}{11},
  \bibinfo{pages}{4111}.
\bibitem[{Currie(2023)}]{currie2023academic}
\bibinfo{author}{Currie, G.M.}, \bibinfo{year}{2023}.
\newblock \bibinfo{title}{Academic integrity and artificial intelligence: is
  chatgpt hype, hero or heresy?}, in: \bibinfo{booktitle}{Seminars in Nuclear
  Medicine}, \bibinfo{organization}{Elsevier}.
\bibitem[{Deriba et~al.(2023)Deriba, Sanusi and Sunday}]{FitsumEnhancing}
\bibinfo{author}{Deriba, F.G.}, \bibinfo{author}{Sanusi, I.T.},
  \bibinfo{author}{Sunday, A.O.}, \bibinfo{year}{2023}.
\newblock \bibinfo{title}{Enhancing computer programming education using
  chatgpt: A mini review}, in: \bibinfo{booktitle}{In 23rd Koli Calling
  International Conference on Computing Education Research (Koli Calling
  ’23)}, \bibinfo{organization}{ACM}.
\bibitem[{Eva A.M.~van Dis(Feb 22, 2023)}]{Eva}
\bibinfo{author}{Eva A.M.~van Dis, Johan~Bollen, W.Z.R.v.R.C.L.B.N.},
  \bibinfo{year}{Feb 22, 2023}.
\newblock \bibinfo{title}{Chatgpt: Five priorities for research}.
\newblock \URLprefix \url{https://www.nature. com/articles/d41586-023-00288-7}.
\bibitem[{Dwivedi et~al.(2023)Dwivedi, Kshetri, Hughes, Slade, Jeyaraj, Kar,
  Baabdullah, Koohang, Raghavan, Ahuja et~al.}]{dwivedi2023so}
\bibinfo{author}{Dwivedi, Y.K.}, \bibinfo{author}{Kshetri, N.},
  \bibinfo{author}{Hughes, L.}, \bibinfo{author}{Slade, E.L.},
  \bibinfo{author}{Jeyaraj, A.}, \bibinfo{author}{Kar, A.K.},
  \bibinfo{author}{Baabdullah, A.M.}, \bibinfo{author}{Koohang, A.},
  \bibinfo{author}{Raghavan, V.}, \bibinfo{author}{Ahuja, M.}, et~al.,
  \bibinfo{year}{2023}.
\newblock \bibinfo{title}{“so what if chatgpt wrote it?” multidisciplinary
  perspectives on opportunities, challenges and implications of generative
  conversational ai for research, practice and policy}.
\newblock \bibinfo{journal}{International Journal of Information Management}
  \bibinfo{volume}{71}, \bibinfo{pages}{102642}.
\bibitem[{Elkhodr et~al.(2023)Elkhodr, Gide, Wu and Darwish}]{elkhodr2023ict}
\bibinfo{author}{Elkhodr, M.}, \bibinfo{author}{Gide, E.}, \bibinfo{author}{Wu,
  R.}, \bibinfo{author}{Darwish, O.}, \bibinfo{year}{2023}.
\newblock \bibinfo{title}{Ict students’ perceptions towards chatgpt: An
  experimental reflective lab analysis}.
\newblock \bibinfo{journal}{STEM Education} \bibinfo{volume}{3},
  \bibinfo{pages}{70--88}.
\bibitem[{Firat(2023)}]{firat2023chatgpt}
\bibinfo{author}{Firat, M.}, \bibinfo{year}{2023}.
\newblock \bibinfo{title}{What chatgpt means for universities: Perceptions of
  scholars and students}.
\newblock \bibinfo{journal}{Journal of Applied Learning and Teaching}
  \bibinfo{volume}{6}.
\bibitem[{Fuchs(2023)}]{fuchs2023exploring}
\bibinfo{author}{Fuchs, K.}, \bibinfo{year}{2023}.
\newblock \bibinfo{title}{Exploring the opportunities and challenges of nlp
  models in higher education: is chat gpt a blessing or a curse?}, in:
  \bibinfo{booktitle}{Frontiers in Education},
  \bibinfo{organization}{Frontiers}. p. \bibinfo{pages}{1166682}.
\bibitem[{Gottipati et~al.(2023)Gottipati, Shim and
  Shankararaman}]{gottipati2023ai}
\bibinfo{author}{Gottipati, S.}, \bibinfo{author}{Shim, K.J.},
  \bibinfo{author}{Shankararaman, V.}, \bibinfo{year}{2023}.
\newblock \bibinfo{title}{Ai for connectivism learning-undergraduate
  students’ experiences of chatgpt in advanced programming courses} .
\bibitem[{Haleem et~al.(2022)Haleem, Javaid, Qadri and
  Suman}]{haleem2022understanding}
\bibinfo{author}{Haleem, A.}, \bibinfo{author}{Javaid, M.},
  \bibinfo{author}{Qadri, M.A.}, \bibinfo{author}{Suman, R.},
  \bibinfo{year}{2022}.
\newblock \bibinfo{title}{Understanding the role of digital technologies in
  education: A review}.
\newblock \bibinfo{journal}{Sustainable Operations and Computers}
  \bibinfo{volume}{3}, \bibinfo{pages}{275--285}.
\bibitem[{Hamid et~al.(2023)Hamid, Zulkifli, Naimat, Yaacob and
  Ng}]{hamid2023exploratory}
\bibinfo{author}{Hamid, H.}, \bibinfo{author}{Zulkifli, K.},
  \bibinfo{author}{Naimat, F.}, \bibinfo{author}{Yaacob, N.L.C.},
  \bibinfo{author}{Ng, K.W.}, \bibinfo{year}{2023}.
\newblock \bibinfo{title}{Exploratory study on student perception on the use of
  chat ai in process-driven problem-based learning}.
\newblock \bibinfo{journal}{Currents in Pharmacy Teaching and Learning} .
\bibitem[{Houghton et~al.(2013)Houghton, Casey, Shaw and
  Murphy}]{houghton2013rigour}
\bibinfo{author}{Houghton, C.}, \bibinfo{author}{Casey, D.},
  \bibinfo{author}{Shaw, D.}, \bibinfo{author}{Murphy, K.},
  \bibinfo{year}{2013}.
\newblock \bibinfo{title}{Rigour in qualitative case-study research}.
\newblock \bibinfo{journal}{Nurse researcher} \bibinfo{volume}{20}.
\bibitem[{Joshi et~al.(2023)Joshi, Budhiraja, Dev, Kadia, Ataullah, Mitra,
  Kumar and Akolekar}]{joshi2023chatgpt}
\bibinfo{author}{Joshi, I.}, \bibinfo{author}{Budhiraja, R.},
  \bibinfo{author}{Dev, H.}, \bibinfo{author}{Kadia, J.},
  \bibinfo{author}{Ataullah, M.O.}, \bibinfo{author}{Mitra, S.},
  \bibinfo{author}{Kumar, D.}, \bibinfo{author}{Akolekar, H.D.},
  \bibinfo{year}{2023}.
\newblock \bibinfo{title}{Chatgpt--a blessing or a curse for undergraduate
  computer science students and instructors?}
\newblock \bibinfo{journal}{arXiv preprint arXiv:2304.14993} .
\bibitem[{Kasneci et~al.(2023)Kasneci, Se{\ss}ler, K{\"u}chemann, Bannert,
  Dementieva, Fischer, Gasser, Groh, G{\"u}nnemann, H{\"u}llermeier
  et~al.}]{kasneci2023chatgpt}
\bibinfo{author}{Kasneci, E.}, \bibinfo{author}{Se{\ss}ler, K.},
  \bibinfo{author}{K{\"u}chemann, S.}, \bibinfo{author}{Bannert, M.},
  \bibinfo{author}{Dementieva, D.}, \bibinfo{author}{Fischer, F.},
  \bibinfo{author}{Gasser, U.}, \bibinfo{author}{Groh, G.},
  \bibinfo{author}{G{\"u}nnemann, S.}, \bibinfo{author}{H{\"u}llermeier, E.},
  et~al., \bibinfo{year}{2023}.
\newblock \bibinfo{title}{Chatgpt for good? on opportunities and challenges of
  large language models for education}.
\newblock \bibinfo{journal}{Learning and individual differences}
  \bibinfo{volume}{103}, \bibinfo{pages}{102274}.
\bibitem[{Khoury et~al.(2023)Khoury, Avila, Brunelle and
  Camara}]{khoury2023secure}
\bibinfo{author}{Khoury, R.}, \bibinfo{author}{Avila, A.R.},
  \bibinfo{author}{Brunelle, J.}, \bibinfo{author}{Camara, B.M.},
  \bibinfo{year}{2023}.
\newblock \bibinfo{title}{How secure is code generated by chatgpt?}
\newblock \bibinfo{journal}{arXiv preprint arXiv:2304.09655} .
\bibitem[{Kohnke et~al.(2023)Kohnke, Moorhouse and Zou}]{kohnke2023chatgpt}
\bibinfo{author}{Kohnke, L.}, \bibinfo{author}{Moorhouse, B.L.},
  \bibinfo{author}{Zou, D.}, \bibinfo{year}{2023}.
\newblock \bibinfo{title}{Chatgpt for language teaching and learning}.
\newblock \bibinfo{journal}{RELC Journal} , \bibinfo{pages}{00336882231162868}.
\bibitem[{Leinonen et~al.(2023)Leinonen, Denny, MacNeil, Sarsa, Bernstein, Kim,
  Tran and Hellas}]{leinonen2023comparing}
\bibinfo{author}{Leinonen, J.}, \bibinfo{author}{Denny, P.},
  \bibinfo{author}{MacNeil, S.}, \bibinfo{author}{Sarsa, S.},
  \bibinfo{author}{Bernstein, S.}, \bibinfo{author}{Kim, J.},
  \bibinfo{author}{Tran, A.}, \bibinfo{author}{Hellas, A.},
  \bibinfo{year}{2023}.
\newblock \bibinfo{title}{Comparing code explanations created by students and
  large language models}.
\newblock \bibinfo{journal}{arXiv preprint arXiv:2304.03938} .
\bibitem[{Ngo(2023)}]{ngo2023perception}
\bibinfo{author}{Ngo, T.T.A.}, \bibinfo{year}{2023}.
\newblock \bibinfo{title}{The perception by university students of the use of
  chatgpt in education}.
\newblock \bibinfo{journal}{International Journal of Emerging Technologies in
  Learning (Online)} \bibinfo{volume}{18}, \bibinfo{pages}{4}.
\bibitem[{Obaido et~al.(2023)Obaido, Agbo, Alvarado and
  Oyelere}]{obaido2023analysis}
\bibinfo{author}{Obaido, G.}, \bibinfo{author}{Agbo, F.J.},
  \bibinfo{author}{Alvarado, C.}, \bibinfo{author}{Oyelere, S.S.},
  \bibinfo{year}{2023}.
\newblock \bibinfo{title}{Analysis of attrition studies within the computer
  sciences}.
\newblock \bibinfo{journal}{IEEE Access} .
\bibitem[{Opara et~al.(2023)Opara, Mfon-Ette~Theresa and
  Aduke}]{opara2023chatgpt}
\bibinfo{author}{Opara, E.}, \bibinfo{author}{Mfon-Ette~Theresa, A.},
  \bibinfo{author}{Aduke, T.C.}, \bibinfo{year}{2023}.
\newblock \bibinfo{title}{Chatgpt for teaching, learning and research:
  Prospects and challenges}.
\newblock \bibinfo{journal}{Opara Emmanuel Chinonso, Adalikwu Mfon-Ette
  Theresa, Tolorunleke Caroline Aduke (2023). ChatGPT for Teaching, Learning
  and Research: Prospects and Challenges. Glob Acad J Humanit Soc Sci}
  \bibinfo{volume}{5}.
\bibitem[{OpenAI(April 25, 2023)}]{OpenAI}
\bibinfo{author}{OpenAI}, \bibinfo{year}{April 25, 2023}.
\newblock \bibinfo{title}{New ways to manage your data in chatgpt}.
\newblock \URLprefix
  \url{https://openai.com/blog/new-ways-to-manage-your-data-in-chatgpt}.
\bibitem[{Ouh et~al.(2023)Ouh, Gan, Shim and Wlodkowski}]{ouh2023chatgpt}
\bibinfo{author}{Ouh, E.L.}, \bibinfo{author}{Gan, B.K.S.},
  \bibinfo{author}{Shim, K.J.}, \bibinfo{author}{Wlodkowski, S.},
  \bibinfo{year}{2023}.
\newblock \bibinfo{title}{Chatgpt, can you generate solutions for my coding
  exercises? an evaluation on its effectiveness in an undergraduate java
  programming course}.
\newblock \bibinfo{journal}{arXiv preprint arXiv:2305.13680} .
\bibitem[{Owens~Brian(Feb 20, 2023)}]{Owens}
\bibinfo{author}{Owens~Brian, N.}, \bibinfo{year}{Feb 20, 2023}.
\newblock \bibinfo{title}{How nature readers are using chatgpt}.
\newblock \URLprefix \url{https://www.nature. com/articles/d41586-023-00500-8}.
\bibitem[{Rahman and Watanobe(2023)}]{rahman2023chatgpt}
\bibinfo{author}{Rahman, M.M.}, \bibinfo{author}{Watanobe, Y.},
  \bibinfo{year}{2023}.
\newblock \bibinfo{title}{Chatgpt for education and research: Opportunities,
  threats, and strategies}.
\newblock \bibinfo{journal}{Applied Sciences} \bibinfo{volume}{13},
  \bibinfo{pages}{5783}.
\bibitem[{Sanusi et~al.(2023)Sanusi, Ayanwale and
  Chiu}]{sanusi2023investigating}
\bibinfo{author}{Sanusi, I.T.}, \bibinfo{author}{Ayanwale, M.A.},
  \bibinfo{author}{Chiu, T.K.}, \bibinfo{year}{2023}.
\newblock \bibinfo{title}{Investigating the moderating effects of social good
  and confidence on teachers' intention to prepare school students for
  artificial intelligence education}.
\newblock \bibinfo{journal}{Education and Information Technologies} ,
  \bibinfo{pages}{1--23}.
\bibitem[{Sanusi et~al.(2022)Sanusi, Olaleye, Oyelere and
  Dixon}]{sanusi2022investigating}
\bibinfo{author}{Sanusi, I.T.}, \bibinfo{author}{Olaleye, S.A.},
  \bibinfo{author}{Oyelere, S.S.}, \bibinfo{author}{Dixon, R.A.},
  \bibinfo{year}{2022}.
\newblock \bibinfo{title}{Investigating learners’ competencies for artificial
  intelligence education in an african k-12 setting}.
\newblock \bibinfo{journal}{Computers and Education Open} \bibinfo{volume}{3},
  \bibinfo{pages}{100083}.
\bibitem[{Shoufan(2023)}]{shoufan2023exploring}
\bibinfo{author}{Shoufan, A.}, \bibinfo{year}{2023}.
\newblock \bibinfo{title}{Exploring students’ perceptions of chatgpt:
  Thematic analysis and follow-up survey}.
\newblock \bibinfo{journal}{IEEE Access} .
\bibitem[{Sok and Heng(2023)}]{sok2023chatgpt}
\bibinfo{author}{Sok, S.}, \bibinfo{author}{Heng, K.}, \bibinfo{year}{2023}.
\newblock \bibinfo{title}{Chatgpt for education and research: A review of
  benefits and risks}.
\newblock \bibinfo{journal}{Available at SSRN 4378735} .
\bibitem[{Surameery and Shakor(2023)}]{surameery2023use}
\bibinfo{author}{Surameery, N.M.S.}, \bibinfo{author}{Shakor, M.Y.},
  \bibinfo{year}{2023}.
\newblock \bibinfo{title}{Use chat gpt to solve programming bugs}.
\newblock \bibinfo{journal}{International Journal of Information Technology \&
  Computer Engineering (IJITC) ISSN: 2455-5290} \bibinfo{volume}{3},
  \bibinfo{pages}{17--22}.
\bibitem[{UNESCO(Sept 8, 2023)}]{UNESCO}
\bibinfo{author}{UNESCO}, \bibinfo{year}{Sept 8, 2023}.
\newblock \bibinfo{title}{Guidance for generative ai in education and
  research}.
\newblock \URLprefix \url{https://unesdoc.unesco.org/ark:/48223/pf0000386693}.
\bibitem[{Vaughan~Connolly(April 5, 2023)}]{Vaughan}
\bibinfo{author}{Vaughan~Connolly, Steve~Watson, U.o.C.}, \bibinfo{year}{April
  5, 2023}.
\newblock \bibinfo{title}{Chatgpt (we need to talk)}.
\newblock \URLprefix \url{https://www.cam.ac.uk/stories/ChatGPT-and-education}.
\bibitem[{Woithe and Filipec(2023)}]{woithe2023understanding}
\bibinfo{author}{Woithe, J.}, \bibinfo{author}{Filipec, O.},
  \bibinfo{year}{2023}.
\newblock \bibinfo{title}{Understanding the adoption, perception, and learning
  impact of chatgpt in higher education: A qualitative exploratory case study
  analyzing students’ perspectives and experiences with the ai-based large
  language model}.
\bibitem[{Wu et~al.(2023)Wu, Duan and Ni}]{wu2023unveiling}
\bibinfo{author}{Wu, X.}, \bibinfo{author}{Duan, R.}, \bibinfo{author}{Ni, J.},
  \bibinfo{year}{2023}.
\newblock \bibinfo{title}{Unveiling security, privacy, and ethical concerns of
  chatgpt}.
\newblock \bibinfo{journal}{arXiv preprint arXiv:2307.14192} .
\bibitem[{Xia et~al.(2022)Xia, Chiu, Lee, Sanusi, Dai and Chai}]{xia2022self}
\bibinfo{author}{Xia, Q.}, \bibinfo{author}{Chiu, T.K.}, \bibinfo{author}{Lee,
  M.}, \bibinfo{author}{Sanusi, I.T.}, \bibinfo{author}{Dai, Y.},
  \bibinfo{author}{Chai, C.S.}, \bibinfo{year}{2022}.
\newblock \bibinfo{title}{A self-determination theory (sdt) design approach for
  inclusive and diverse artificial intelligence (ai) education}.
\newblock \bibinfo{journal}{Computers \& Education} \bibinfo{volume}{189},
  \bibinfo{pages}{104582}.
\bibitem[{Yilmaz and Yilmaz(2023)}]{yilmaz2023augmented}
\bibinfo{author}{Yilmaz, R.}, \bibinfo{author}{Yilmaz, F.G.K.},
  \bibinfo{year}{2023}.
\newblock \bibinfo{title}{Augmented intelligence in programming learning:
  Examining student views on the use of chatgpt for programming learning}.
\newblock \bibinfo{journal}{Computers in Human Behavior: Artificial Humans}
  \bibinfo{volume}{1}, \bibinfo{pages}{100005}.
\bibitem[{Yunusa et~al.(2021)Yunusa, Sanusi, Dada, Oyelere, Agbo, Obaido and
  Aruleba}]{yunusa2021impact}
\bibinfo{author}{Yunusa, A.A.}, \bibinfo{author}{Sanusi, I.T.},
  \bibinfo{author}{Dada, O.A.}, \bibinfo{author}{Oyelere, S.S.},
  \bibinfo{author}{Agbo, F.J.}, \bibinfo{author}{Obaido, G.},
  \bibinfo{author}{Aruleba, K.}, \bibinfo{year}{2021}.
\newblock \bibinfo{title}{The impact of the covid-19 pandemic on higher
  education in nigeria: university lecturers’ perspectives}.
\newblock \bibinfo{journal}{IjEDict-International Journal of Education and
  Development Using Information and Communication Technology}
  \bibinfo{volume}{17}, \bibinfo{pages}{43--66}.

\end{thebibliography}


\end{document}